\documentclass[conference]{IEEEtran}
\IEEEoverridecommandlockouts
\usepackage{cite}
\usepackage{amsmath,amssymb,amsfonts}
\usepackage{algorithmic}
\usepackage{graphicx}
\usepackage{textcomp}
\usepackage{xcolor}

\def\BibTeX{{\rm B\kern-.05em{\sc i\kern-.025em b}\kern-.08em
    T\kern-.1667em\lower.7ex\hbox{E}\kern-.125emX}}
\begin{document}

\title{Nonlinear symbols combining for Power Amplifier-distorted OFDM signal reception \\
\thanks{This research was funded by the Polish National Science Centre, projects no. 2021/41/B/ST7/00136 and 2023/05/Y/ST7/00002. For the purpose of Open Access, the authors applied a CC-BY public copyright license to any Author Accepted Manuscript (AAM) version arising from this submission. }
}

\author{\IEEEauthorblockN{Pawel Kryszkiewicz}
\IEEEauthorblockA{\textit{Institute of Radiocommunications} \\
\textit{Poznan University of Technology}\\
Poznan, POLAND \\
pawel.kryszkiewicz@put.poznan.pl}
\and
\IEEEauthorblockN{Hanna Bogucka}
\IEEEauthorblockA{\textit{Institute of Radiocommunications} \\
\textit{Poznan University of Technology}\\
Poznan, POLAND \\
hanna.bogucka@put.poznan.pl}
}

\maketitle

\begin{abstract}
Nonlinear distortion of a multicarrier signal by a transmitter Power Amplifier (PA) can be a serious problem when designing new highly energy-efficient wireless systems. Although the performance of standard reception algorithms is seriously deteriorated by the nonlinear distortion, the more advanced solutions allow the utilization of additional frequency diversity caused by nonlinear PA. However, while most of the advanced receivers are decision-aided, their gains are observed mostly in a relatively low Bit Error Rate (BER) region, not targeted by adaptive Modulation Coding Schemes utilizing Forward Error Correction (FEC). In this paper, a non-decision-aided Higher-Order Combining (HOC) reception scheme is proposed. While the analytical formulas for finding symbols combining coefficients are not known, machine learning is used for deriving them. The simulation results show an improved BER performance with respect to a standard reception and one of the established decision-aided receivers. However, as HOC has computational complexity that increases rapidly with the number of subcarriers utilized, more studies are needed to apply it in a wideband system. 
\end{abstract}


\section{Introduction}
The constantly increasing demand for wireless connectivity increases requirements for wireless transceivers, e.g., their bandwidth and throughput. At the same time, the cost of hardware components is expected to drop together with an expected reduction in the cost of operation measured mainly by the costs of energy. To meet these two opposite goals, new degrees of freedom in the design of wireless terminals are to be taken advantage of. Although some solutions, such as massive MIMO, are commonly known\cite{Bjornson_2018_infinite_capacity}, inclusion of radio front-end characteristics in the design of transmission and reception algorithms or resource allocation can provide significant gains in terms of both the Spectral Efficiency (SE) and the Energy Efficiency (EE) \cite{Fettweis_DirtyRF_2005,Kryszkiewicz_Battery_2023}. The issue is that every characteristic of the wireless front-end is non-linear if pushed into a high EE region. Typically, the front ends are modeled as transparent to the incoming signals, and a multipath channel and additive white noise are considered the main sources of distortion. A realistic front end introduces some distortion that, if modeled realistically, makes the transceiver design much more difficult.

The primary part of energy consumption and non-linear distortions is typically associated with the Power Amplifier (PA). When transmitting Orthogonal Frequency Division Multiplexing (OFDM) waveform, e.g., in 5G New Radio, characterized by large complex envelope fluctuations \cite{Wei_2010_dist_OFDM}, non-linear PA introduces significant nonlinear distortion. This effect can be reduced by signal processing decreasing the peak-to-average power ratio \cite{Kryszkiewicz_TR_2017, Hu_TR_Nonlinearity_2015,Bogucka_2006}. Recently, it has been observed that nonlinear distortion can be controlled, by optimizing the PA operating point, that maximizes SE or EE \cite{Tavares_2016_IBO,Kryszkiewicz_Battery_2023}.  

The last option, which is the main focus of this paper, is the use of PA awareness for advanced OFDM signal reception. While in the standard reception the nonlinear distortion can be treated similarly as white noise, the distortion signal depends on the wanted signal. It has been shown in \cite{Guerreiro_2013_optimum_RX} that non-linearly distorted OFDM can even achieve higher throughput than a linear system thanks to the additional frequency diversity obtained by non-linear processing. However, an optimal Maximal Likelihood (ML) receiver (RX) has prohibitively high computational complexity. Several decision-directed solutions have lower computational complexity, e.g., the Clipping Noise Cancellation (CNC) algorithm \cite{ochiai_cnc_and_dar_eval,Wachowiak_2023} or ones that require multiple iterations of symbol detection and distortion reconstruction or a belief propagation reception algorithm \cite{belief_propagation_rx_ofdm}. Recently, solutions utilizing neural networks have been proposed for this purpose \cite{1DTRNet, HybridDeepRx}. However, all these solutions focus mainly on improving reception performance for a low BER region, e.g., below $10^{-2}$. In this case, most of the transmitted symbols are received correctly, and thus, decision-directed solutions can provide significant gains. However, because typically OFDM transmission is protected by some strong channel coding, e.g. LDPC coding in 5G NR, the main gain is required for a much higher uncoded BER region.

This paper proposes a Higher Order Combining (HOC) scheme to allow for non-decision-directed reception of an OFDM signal under severe nonlinear PA distortion. The proposed scheme is capable of using the frequency diversity resulting from non-linear PA processing. While the analytical combining coefficients are not known and are difficult to obtain, Machine Learning (ML) is used to derive them based on the proposed physics-based model. Most interestingly, the initial analysis reveals that only specific higher-order combining coefficients are required for the reception. While the computational complexity of both learning and inference increases rapidly with the number of subcarriers, a reduced complexity solution is also proposed. Computer simulations under the fading channel confirmed the superiority of the proposed scheme over both standard and decision-aided reception in the BER range of interest. The considered OFDM system model with nonlinear PA is shown in Sec. \ref{sec_System_model}, and the proposed HOC scheme is derived in Sec. \ref{sec_proposed_method}. The simulation results are presented in Sec. \ref{sec_Simulation}, and the paper is concluded in Sec. \ref{sec_Conclusions}.

\section{System Model}
\label{sec_System_model}
\begin{figure}[!t]
\centering
\includegraphics[width=0.9\columnwidth]{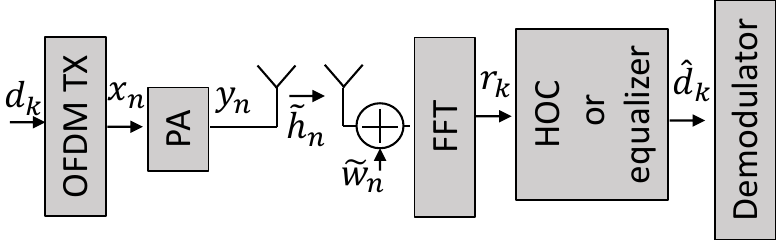}
\caption{System model}
\label{fig:scheme}
\end{figure}

The considered OFDM link is shown in Fig. \ref{fig:scheme}. The transmitter (TX) is composed of an $N$-point Inverse Fast Fourier Transform (IFFT) block with the $n$-th sample on its output defined as
\begin{equation}
    x_n=\frac{1}{\sqrt{N}}\sum_{k=0}^{N_{U}-1} d_k e^{j 2 \pi \frac{n I_k}{N}},
    \label{eq_IFFT}
\end{equation}
where $d_k$ is the complex symbol, typically from the Quadrature Amplitude Modulation (QAM) or Phase Shift Keying (PSK) constellations, modulating $I_k$-th subcarrier, and $n\in \{ -N_{\mathrm{CP}},..., N-1\}$ for $N_{\mathrm{CP}}$ being the length of a cyclic prefix (in samples). The indices of all $N_{\mathrm{U}}$ occupied subcarriers are denoted by a vector $I$ composed of unique
elements from a total set of available subcarriers $\{ -N/2, ...., N/2-1\}$. 
The unoccupied subcarriers are left to allow for digital-analog processing, reducing the aliasing effect. 
The mean power of a single, time-domain OFDM sample can be denoted as
\begin{equation}
    \sigma^2=\mathbb{E} \left[ |x_n|^2\right].
    \label{eq_power}
\end{equation}
The digital signal represented by samples $\{x_n\}$ undergoes Digital-Analog Conversion (DAC) and upconversion before being transmitted by PA. Here, an equivalent baseband discrete-time model is used, denoted by a transfer function $\Gamma(\cdot)$, i.e., the PA output equals
    $y_n=\Gamma \left( x_n \right)$.
While there are multiple PA models available\cite{Gharaibeh_distortion_book}, here a Rapp model is used:
\begin{equation}
    y_n = \frac{Gx_n}{(1+\frac{|G x_n|^{2p}}{P_{\mathrm{max}}^p})^{\frac{1}{2p}}},
\end{equation}
for which $G$ is the linear signal gain, $P_{\mathrm{max}}$ denotes PA saturation power and $p$ is the smoothing factor. Note that $p=2$ is reported to be close to practical solid-state power amplifiers \cite{Ochiai_2013_PA_efficiency}, and $p \to \infty$ models a soft-limiter (or clipper). Moreover, in practical systems, digital predistortion of the PA input signal is typically performed \cite{Wyglinski_predistortion_2016}. In such a case, the transfer function contains the combined characteristics of a predistorter and the PA with a higher $p$ value than for the PA alone.
An important parameter characterizing the operating point of the PA is Input Back-Off (IBO) specified as
\begin{equation}
    \gamma=\frac{P_{\mathrm{MAX}}}{\sigma^2}.
\end{equation}
Additionally, for the PA transfer function, a memoryless polynomial model \cite{Gharaibeh_distortion_book} can be considered, motivating the reception method proposed in the later part of this paper. It is defined as
\begin{equation}
    y_n=\sum_{q=1}^{Q} a_q x_n |x_n|^{2(q-1)},
    \label{eq_polynomial}
\end{equation}
where $a_q$ denotes $q$-th monomial coefficient, and $Q$ is the number of considered monomials. Most importantly, for the baseband PA modeling, only odd order monomials are used.  

While the OFDM waveform, for a sufficiently high number of occupied subcarriers, is complex-Gaussian distributed \cite{Wei_2010_dist_OFDM}, a Bussgang decomposition can be applied to partition the PA-output signal as:
\begin{equation}
y_n=\alpha x_n +\tilde{q}^{\mathrm{PA}}_{n},
\label{eq_busggang}
\end{equation}
where
\begin{equation}
    \alpha=\frac{\mathbb{E}\left[y_n x_n^* \right]}{\mathbb{E}\left[x_n x_n^* \right]}
    \label{eq_lambda_def}
\end{equation}
can be calculated numerically for the Rapp model  \cite{Kryszkiewicz_Battery_2023}, and $\tilde{q}^{\mathrm{PA}}_{n}$ is uncorrelated with $x_n$ distortion signal.

The signal passes through a multipath channel of impulse response $h_l$ for $l\in \{ 0,..., L-1\}$. After the addition of additive white noise $\tilde{w}_n$ the $n$-th sample of the received signal equals
\begin{equation}
    \tilde{r}_n=\sum_{l=0}^{L-1}\tilde{h}_l y_{n-l} + \tilde{w}_n.
    \label{eq_time_channel}
\end{equation}
The received signal at $k$-th used subcarrier is obtained by Fast Fourier Transform (FFT) that after substitution of (\ref{eq_time_channel}) gives 
\begin{align}
    \label{eq_RX_signal_initial}
    r_k&=\frac{1}{\sqrt{N}}\sum_{n=0}^{N-1} \tilde{r}_n e^{-j 2 \pi \frac{n I_k}{N}}
    \\ \nonumber
        &=\underbrace{\sum_{l=0}^{L-1}h_l e^{-j 2 \pi \frac{l I_k}{N}}}_{h_{k}} \underbrace{\frac{1}{\sqrt{N}} \sum_{n=0}^{N-1}y_{n-l} e^{-j 2 \pi \frac{(n-l) I_k}{N}}}_{\textrm{DFT of $y_n$ by its periodcity with CP} }
        + w_k,
\end{align}
where $w_k$ and $h_{k}$ are AWGN noise sample and the complex channel response at $k$-th used subcarrier, respectively. 

\subsection{Standard reception}
\label{sec_standard_RX}
If (\ref{eq_busggang}) is used as the PA model, the signal $r_k$ can be further simplified to
\begin{equation}
    r_k=h_k \alpha d_k +h_k {q}_k^{\mathrm{PA}} +w_k,
    \label{eq_r_k_dinkelbach}
\end{equation}
where ${q}_k^{\mathrm{PA}}$ is the result of DFT processing of $\tilde{q}_n^{\mathrm{PA}}$.
If the nonlinear distortion is treated as part of the noise\cite{ochiai_cnc_and_dar_eval}, the Zero Forcing (ZF) estimate of $d_k$ equals
\begin{equation}
    \hat{d}^{\mathrm{standard}}_k=d_k +\frac{\tilde{q}_k^{\mathrm{PA}}}{\alpha} +\frac{w_k}{h_k \alpha}.
    \label{eq_standard_RX}
\end{equation}

\section{PA-Aware Higher Order Combining (HOC)}
\label{sec_proposed_method}
Another approach to signal reception is to consider (\ref{eq_polynomial}) as the PA model, that after substitution to  (\ref{eq_RX_signal_initial}) and using (\ref{eq_IFFT}) allows to define the received frequency domain signal as:
\begin{subequations}
\begin{align}
    &r_k=h_k a_1 d_k 
    \label{eq_linear}
    \\& +h_k a_2\!\!\!\!\!\!\!\sum_{k1,k2,k3=0}^{N_{\mathrm{U}}} \!\!\!\!\!\! d_{k1} d_{k2} d_{k3}^{*} \frac{1}{N} \sum_{n=0}^{N-1} e^{j 2 \pi \frac{n \left(I_{k1}+I_{k2}-I_{k3}-I_{k}\right)}{N}}
    \label{eq_IMD3}
   \\&
   +h_k a_3  ...
   \\&
   = h_k a_1 d_k +\underbrace{h_k a_2 \!\!\!\!\!\sum_{\substack{k1,k2,k3: \\ I_{k1}+I_{k2}-I_{k3}-I_{k}=0}}^{N_{\mathrm{U}}}\!\!\!\!\! d_{k1} d_{k2} d_{k3}^{*}}_{r_k^{\mathrm{IMD3}}} +h_k a_3...,
   \label{eq_RX_raw_symbols}
\end{align}
\end{subequations}
where only linear amplification term in (\ref{eq_linear}) and 3rd order intermodulations (IMD3) in (\ref{eq_IMD3}) are fully represented. In the derivation of IMD3 signal, a formula for geometric progression can be used that is non-zero only for $I_{k1}+I_{k2}-I_{k3}-I_{k}=0$ while subcarrier indices are integers. Consequently, if there are multiple subcarriers, the symbols modulating them at TX are distributed over multiple RX subcarriers. For example, in an OFDM system that utilizes three subcarriers of indices $I=\{0,1,2\}$: 
\begin{equation}
    r_0^{\mathrm{IMD3}}=h_0 a_2 d_0 (|d_0|^2+2|d_1|^2+|d_2|^2)+d_1 d_1 d_2^{*}
\end{equation}
depends on $d_0$, $d_1$ and $d_2$. While the energy of each wanted symbol $d_k$ is spread over multiple received symbols $r_k$ we can state a hypothesis that by a proper combining of $r_k$ symbols, the accuracy of $d_k$ estimate can be higher. However, because of the nonlinear nature of the phenomenon, higher-order combining should be used. This hypothesis has been tested by defining initially the 3rd order combining function:
\begin{align}
\nonumber
    \hat{d}_{k}&=\sum_{k1=0}^{N_{\mathrm{U}}}b^{(1)}_{k,k1} r_{k1} + \sum_{k1=0}^{N_{\mathrm{U}}}b^{(2)}_{k,k1} r_{k1}^{*} 
    \\&\nonumber
    +\sum_{k1=0,k2=k1}^{N_{\mathrm{U}}}b^{(3)}_{k,k1,k2} r_{k1}r_{k2}+\sum_{k1,k2=0}^{N_{\mathrm{U}}}b^{(4)}_{k,k1,k2} r_{k1}^{*}r_{k2}
    \\&\nonumber
    +\sum_{\substack{k1=0,\\k2=k1}}^{N_{\mathrm{U}}}b^{(5)}_{k,k1,k2} r_{k1}^{*}r_{k2}^{*}+\sum_{\substack{k1=0,k2=k1,\\k3=k2}}^{N_{\mathrm{U}}}b^{(6)}_{k,k1,k2,k3} r_{k1}r_{k2}r_{k3}
    \\&\nonumber
    +\!\!\!\!\!\!\sum_{\substack{k1,k2=0,\\k3=k2}}^{N_{\mathrm{U}}}\!\!\!\!\!\!\!\!b^{(7)}_{k,k1,k2,k3} r_{k1}^{*}r_{k2}r_{k3}+\!\!\!\!\!\!\!\!\sum_{\substack{k3=0,k2=0,\\k1=k2}}^{N_{\mathrm{U}}}\!\!\!\!\!\!\!\!b^{(8)}_{k,k1,k2,k3} r_{k1}^{*}r_{k2}^{*}r_{k3}
    \\&
    +\sum_{k1=0,k2=k1,k3=k2}^{N_{\mathrm{U}}}b^{(9)}_{k,k1,k2,k3} r_{k1}^{*}r_{k2}^{*}r_{k3}^{*},
    \label{eq_full_combining}
\end{align}
with $\mathbf{b}_k=[b^{(1)}_{k,0},b^{(1)}_{k,1},...,b^{(2)}_{k,0},...,b^{(9)}_{k,N_{\mathrm{U}},N_{\mathrm{U}},N_{\mathrm{U}}}]^{\mathrm{T}}$ being vectorized combining coefficients for the $k$-th subcarrier. The optimization can be carried out to find $\mathbf{b}_k$ such that the mean squared error between the estimated symbols and the transmitted symbols is minimized:
\begin{equation}
    \min_{\mathbf{b}_k} \mathbb{E} \left[\left|\hat{d}_{k} -d_k \right|^{2}\right],
\end{equation}
independently for each subcarrier.
While finding these coefficients analytically can be difficult, ML based on appropriate training datasets can be performed. Frequency-domain data-driven learning has been already used for OFDM in the past, e.g., \cite{Bogucka2000} for echo cancellation in a full-duplex transceiver. The dataset can be obtained by transmitting thousands of random OFDM symbols over a given PA (or its digital twin model) and observing the received symbols $r_k$. The problem can be solved by vectorizing the problem as 
\begin{equation}
    \min_{\mathbf{b}_k} \left\|\mathbf{R}^{\mathrm{IMD3}}\mathbf{b}_k -\mathbf{d}_k \right\|^{2},
\end{equation}
where $\mathbf{R}^{\mathrm{IMD3}}$ is a matrix with each row representing a separate test symbol and each column denoting a single monomial from (\ref{eq_full_combining}) and $\mathbf{d}_k$ is a vector of the all symbols modulating the $k$-th subcarrier. 
The solution can be obtained by Moore-Penrose pseudoinverse as
\begin{equation}
    \mathbf{\widehat{b}}_k=\mathbf{R}^{\mathrm{IMD3}+}\mathbf{d}_k.
    \label{eq_pseudoinverse_IMD3}
\end{equation}
Most interestingly, initial learning in a system without white noise, under Rapp-modeled PA, revealed that only a limited set of combining coefficients has any influence on the estimated received symbols. Based on multiple tested use cases (with varying numbers and indices of subcarriers), the formula (\ref{eq_full_combining}) can be simplified to 
\begin{align}
    \hat{d}_{k}^{\mathrm{IMD3}}&=c^{(1)}_{k} r_{k}+\!\!\!\!\!\!\sum_{\substack{k1=0,k2=k1,k3=0: \\ I_{k1}+I_{k2}-I_{k3}-I_{k}=0}}^{N_{\mathrm{U}}} \!\!\!\!\!\! c^{(2)}_{k,k1,k2,k3} r_{k1}r_{k2}r_{k3}^{*}.
    \label{eq_adv_RX_IMD3}
\end{align}
The above formula shows that the HOC applied to estimate $d_k$ takes into account linear term $r_k$, similarly as in the standard reception (\ref{eq_standard_RX}), as well as the same third-order terms as used in (\ref{eq_RX_raw_symbols}), although in (\ref{eq_RX_raw_symbols}) symbols $d_k$, not $r_k$ as in (\ref{eq_adv_RX_IMD3}). With this observation in mind, additionally, the 5th order terms can be used in the following combining scheme: 
\begin{align}
    \hat{d}_{k}^{\mathrm{IMD5}}&=c^{(1)}_{k} r_{k}+\!\!\!\!\!\!\sum_{\substack{k1=0,k2=k1,k3=0: \\ I_{k1}+I_{k2}-I_{k3}-I_{k}=0}}^{N_{\mathrm{U}}} \!\!\!\!\!\! c^{(2)}_{k,k1,k2,k3} r_{k1}r_{k2}r_{k3}^{*} 
        \label{eq_adv_RX_IMD5}
    \\&
    +\mkern-30mu  \sum_{\substack{k1=0,k2=k1,k3=k2,k4=0,k5=k4: \\ I_{k1}+I_{k2}+I_{k3}-I_{k4}-I_{k5}-I_{k}=0}}^{N_{\mathrm{U}}} \mkern-50mu c^{(3)}_{k,k1,k2,k3,k4,k5} r_{k1}r_{k2}r_{k3}r_{k4}^{*}r_{k5}^{*}.
    \nonumber
\end{align}
Its solution can be obtained, similarly to (\ref{eq_pseudoinverse_IMD3}), by pseudoinverse as
\begin{equation}
    \mathbf{\widehat{c}}_k=\mathbf{R}^{\mathrm{IMD5}+}\mathbf{d}_k,
    \label{eq_pseudoinverse_IMD5}
\end{equation}
with $\mathbf{c}_k$ being vectorized combining coefficients for $k$-th subcarrier and $\mathbf{R}^{\mathrm{IMD5}}$ being a matrix with each row representing a separate test symbol and each column denoting a single monomial from (\ref{eq_adv_RX_IMD5}).
After the coefficients $\mathbf{\widehat{c}}_k$ are obtained, the estimate of $\hat{d}_{k}^{\mathrm{IMD5}}$ can be obtained by (\ref{eq_adv_RX_IMD5}). 

While the estimation of $\mathbf{\widehat{c}}_k$ coefficients should happen for some training data, e.g., performed in some link's digital twin, the reception happens for a different received symbols matrix, i.e., $\mathbf{R}^{\mathrm{IMD5}}$ changes between learning and inference phases.

Additionally, observe that the proposed HOC uses only the symbols received at the utilized subcarriers. While some of the nonlinear distortion is emitted at unused subcarriers, in the so-called Out Of Band (OOB) region, this can potentially improve the reception performance even further. However, in practical systems, the OOB components can be sometimes filtered out at the TX to reduce distortion to adjacent band systems. Secondly, the wireless channel is not known for these subcarriers, therefore, utilizing them for signal reception would require some advanced channel estimation procedures.    

While the above HOC methods are defined for IMD3 and IMD5, one can ask why not extend it to higher intermodulation (IMD) terms. It can be shown for a soft limiter \cite{lee2014characterization} that depending on the IBO value, typically, a significant power of nonlinear distortion is present in higher-order IMD terms. With the hypothesis that higher-order IMDs can be removed only by higher-order combining terms, their addition can significantly improve reception quality. However, this will result in a significant increase in computational complexity during both the training phase and the inference phase. First, the number of monomials (and coefficients) in (\ref{eq_adv_RX_IMD5}) increases rapidly with the number of subcarriers and IMD orders. For example, for $N_{\mathrm{U}}=6$, there are around 14 IMD3 terms and 100 IMD5 terms per each estimated symbol $\hat{d}_{k}^{\mathrm{IMD5}}$ (although these numbers differ slightly depending on whether the subcarrier is in the middle or at the edge of the occupied band), for $N_{\mathrm{U}}=12$ subcarriers the number of IMD3 and IMD5 terms is around 50 and 1300, respectively. The learning phase needs a much higher number of OFDM symbols for training than the number of combining coefficients to allow for proper noise averaging by an overdetermined linear system estimation by pseudoinverse. Moreover, the computational complexity of the pseudoinverse calculation scales with the number of IMD terms or a number of subcarriers. Finally, the learning stage should be repeated if the wireless channel changes (either its impulse response or Signal to Noise power Ratio (SNR)). 
Therefore, in the next subsection, a reduced complexity HOC algorithm is proposed. 

\subsection{Low-complexity HOC}
The most computationally complex part of the proposed reception method is estimating the coefficients every time the wireless channel or SNR changes. However, a low-complexity (LC) combining scheme, LC-HOC, can be proposed that takes into account only nonlinear distortion, not considering white noise or frequency-specific fading during learning. 
Let us define the received signal at $k$-th subcarrier if observed directly at the TX PA output, based on (\ref{eq_r_k_dinkelbach}) as 
\begin{equation}
    \tilde{r}_k=\alpha d_k + {q}_k^{\mathrm{PA}}.
\end{equation}
The 5th order combining scheme can now be defined as
\begin{align}
    \tilde{d}_{k}^{\mathrm{IMD5}}&=\tilde{c}^{(1)}_{k} \tilde{r}_{k}+\!\!\!\!\!\!\sum_{\substack{k1=0,k2=k1,k3=0: \\ I_{k1}+I_{k2}-I_{k3}-I_{k}=0}}^{N_{\mathrm{U}}} \!\!\!\!\!\! \tilde{c}^{(2)}_{k,k1,k2,k3} \tilde{r}_{k1}\tilde{r}_{k2}\tilde{r}_{k3}^{*} 
        \label{eq_adv_RX_IMD5_simplify}
    \\&
    +\mkern-30mu  \sum_{\substack{k1=0,k2=k1,k3=k2,k4=0,k5=k4: \\ I_{k1}+I_{k2}+I_{k3}-I_{k4}-I_{k5}-I_{k}=0}}^{N_{\mathrm{U}}} \mkern-50mu \tilde{c}^{(3)}_{k,k1,k2,k3,k4,k5} \tilde{r}_{k1}\tilde{r}_{k2}\tilde{r}_{k3}\tilde{r}_{k4}^{*}\tilde{r}_{k5}^{*},
    \nonumber
\end{align}
where $\tilde{c}^{(1)}_{k}$, $\tilde{c}^{(2)}_{k,k1,k2,k3}$ and $\tilde{c}^{(3)}_{k,k1,k2,k3,k4,k5}$ are the new combining coefficients. The coefficients can be obtained by solving the optimization problem
\begin{equation}
    \min_{\tilde{\mathbf{c}}_k} \mathbb{E} \left[\left|\tilde{d}_{k}^{\mathrm{IMD5}} -d_k \right|^{2}\right],
\end{equation}
by using pseudoinverse, similarly to (\ref{eq_pseudoinverse_IMD5}). Observe that this process is independent of the wireless channel changes or noise. Thus, the combining coefficients are specific for a given PA applied at TX, and can be precomputed. 
For symbol detection in RX, the impact of the wireless channel must be removed, e.g., by ZF equalization before applying (\ref{eq_adv_RX_IMD5_simplify}). 
Consequently, symbols $\frac{r_k}{h_k}$ are used instead of $\tilde{r}_{k}$ in (\ref{eq_adv_RX_IMD5_simplify}).  
 
\section{Simulation Results}
\label{sec_Simulation}
The computer simulations are carried out using $N_{\mathrm{U}}=6$ subcarriers modulated with 64-QAM symbols with 64 subcarrier OFDM system. The combining coefficients are obtained as a result of supervised learning,  namely regression based on $10^4$ random OFDM symbols. However, the inference/symbols detection to obtain BER figures is carried out using a separate set of $10^4$ OFDM symbols with independent white noise generated. The independent, identically distributed Rayleigh channel per subcarrier is used with 1000 instances generated (in each instance, $10^4$ OFDM symbols are transmitted). PA is modeled using the Rapp model with a smoothness factor $p=10$ of various IBO. The proposed HOC schemes are configured to use only 1st and 3rd or 1st, 3rd, and 5th order combining terms. These are compared against the reference solution (REF) presented in Sec. \ref{sec_standard_RX} and the Clipping Noise Cancellation (CNC) algorithm \cite{ochiai_cnc_and_dar_eval}, being a decision-aided solution, using 10 iterations. 
    
\begin{figure}[!t]
\centering
\includegraphics[width=0.9\columnwidth]{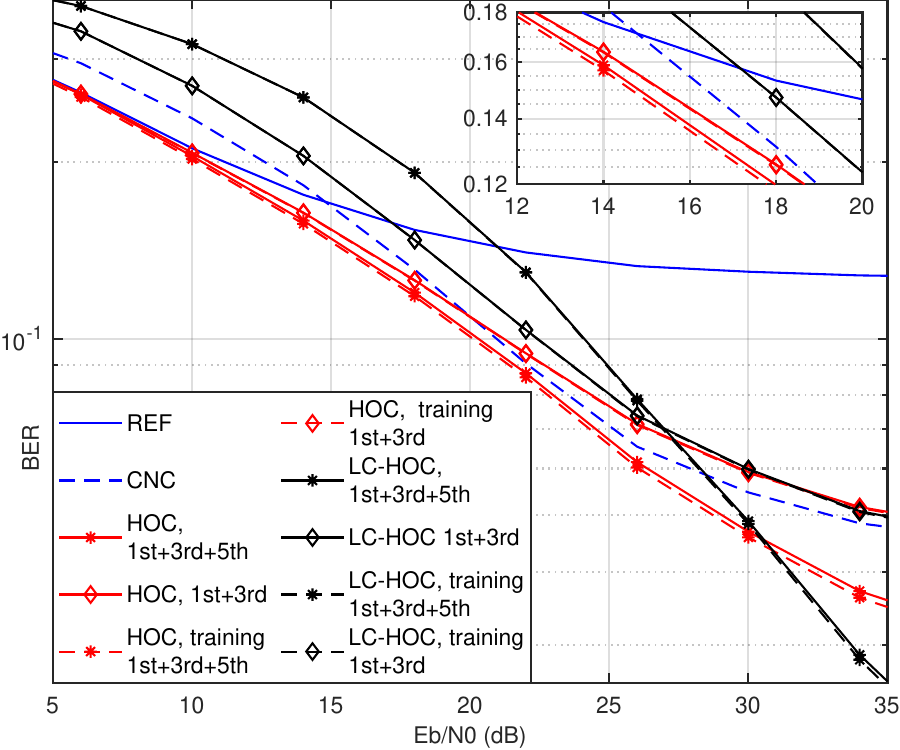}
\caption{BER vs. $E_{\mathrm{b}}/N_0$ for IBO equal to $-4$~dB.}
\label{fig:BER_vs_EBN0}
\end{figure}

BER as a function of energy-per-bit over the noise power spectral density ($E_{\mathrm{b}}/N_0$) is shown in Fig. \ref{fig:BER_vs_EBN0} for IBO of $-4$~dB. This IBO value causes a significant amount of distortion. It is visible that in the low $E_{\mathrm{b}}/N_0$ region, the best performance is obtained by full HOC using 1st+3rd or 1st+3rd+5th combining coefficients. For higher $E_{\mathrm{b}}/N_0$ values, the HOC variant using the 5th order combining coefficients gains in performance. The LC-HOC schemes are outperformed by all other receivers in the low $E_{\mathrm{b}}/N_0$ region, as a result of not considering white noise in the combining-coefficient design.
Additionally, observe that in all the cases, BER obtained for training data and for test (inference) data, result in similar performance, showing that there is no overfitting in this process. 
Most interestingly, the LC-HOC scheme outperforms all other solutions for high $E_{\mathrm{b}}/N_0$. This needs further investigation to understand why the basic HOC is outperformed. In the most important BER region of the FEC-protected system, around 0.15, HOC using 5rd order terms gains around 1.5 dB with respect to the decision-aided CNC RX. 
 
\begin{figure}[!t]
\centering
\includegraphics[width=0.9\columnwidth]{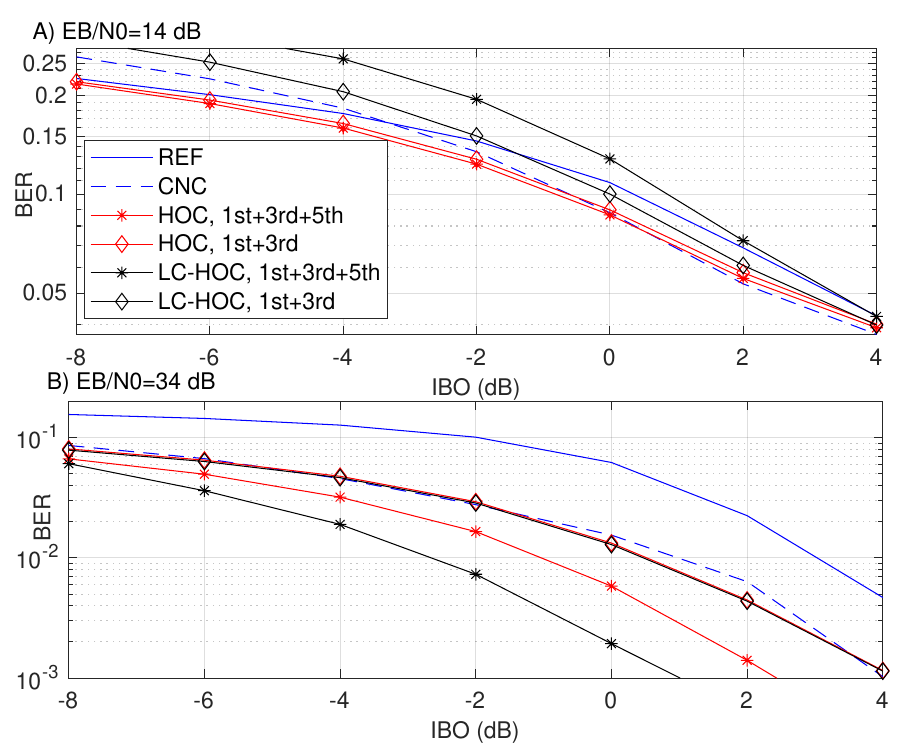}
\caption{BER vs IBO for $E_{\mathrm{b}}/N_0$=14 dB and 34 dB.}
\label{fig:BER_vs_IBO}
\end{figure}

Fig. \ref{fig:BER_vs_IBO} shows BER vs. IBO for two selected $E_{\mathrm{b}}/N_0$ values. For $E_{\mathrm{b}}/N_0=14$ dB (Fig. \ref{fig:BER_vs_IBO}A), HOC outperforms CNC for IBO lower than 0 dB (severe clipping). However, as the clipping becomes less prominent, HOC is outperformed by CNC algorithm. As shown in \cite{Tavares_2016_IBO,Kryszkiewicz_Battery_2023}, even for standard OFDM reception, significant clipping can be an optimal solution under some channel conditions. Most interestingly, under negligible noise power (Fig. \ref{fig:BER_vs_IBO}B), HOC using the 3rd and 5th order terms outperforms CNC in the whole IBO range. It shows that HOC has potential in fighting the reception performance degradation by nonlinear effect but needs some further research in designing its coefficients under high-power noise. 

\section{Conclusions}
\label{sec_Conclusions}
The proposed higher-order combining scheme effectively improves multicarrier signal reception performance under severe nonlinear distortions (caused by TX PA) in a relatively high-BER region, where the decision-aided solutions fail. While the computational complexity increases fast with the number of subcarriers or the combining order, further research is needed to speed up both learning and inference process. Potentially, semi-analytical solutions can be used. Most importantly, a completely new branch for advanced distortion-aware receivers has been created.     

\bibliographystyle{IEEEtran}
\bibliography{bib_PK}
\end{document}